\documentclass[12pt,a4paper]{scrartcl}
\usepackage{graphicx}
\usepackage[T1]{fontenc}
\usepackage[utf8]{inputenc}
\usepackage{lmodern}
\usepackage[stretch=10,shrink=10]{microtype}
\usepackage{hyperref}

\begin{document}

\title{Meandering of liquid rivulets on partially wetting inclines}
\author{Stéphanie Couvreur and Adrian Daerr\footnote{Matière et Systèmes Complexes, UMR 7057, Université Paris Diderot, 75205 Paris cedex 13, France}}  
\date{\normalsize \href{http://arxiv.org/abs/1210.3902}{arXiv:1210.3902 [physics.flu-dyn]}}

\maketitle

\vspace*{-2em}
\begin{abstract}\small
  This fluid dynamics video illustrates recent advances in the
  understanding of the mechanism which causes the sinuous path formed
  by liquid rivulets on partially wetting inclines. The images
  themselves show how a simple lighting set-up using a large Fresnel
  lens can be used to obtain high contrast images of large transparent
  objects.
\end{abstract}

\noindent
Video links\footnote{%
Alternate video links --- arXiv: \url{http://arxiv.org/src/1210.3902/anc}\\
Youtube: \url{https://www.youtube.com/watch?v=vMygH-g0WKU}
}: \href{http://www.msc.univ-paris-diderot.fr/~daerr/gfm2012/Couvreur-Daerr_meandering-rivulets_gfm2012_small.mp4}{$640\times
360$ version, 6MB}
\href{http://www.msc.univ-paris-diderot.fr/~daerr/gfm2012/Couvreur-Daerr_meandering-rivulets_gfm2012.mp4}{$1280\times
720$ version, 34MB}
\medskip

The meandering of liquid rivulets is an ubiquitous phenomenon whose
mechanism has however remained subject to debate\cite{general}. While
the main intervening forces are quickly listed, the wetting hysteresis
found on most partial wetting substrates causes contact line pinning,
which makes modelling difficult. The recent description of a system
exhibiting meandering in total wetting conditions\cite{drenckhan2004}
has led us to measure and calculate the dynamics of the rivulet and to
identify an inertial linear instability, with good agreement between
theory and experiment\cite{daerr2011}.

We revisit meandering in partial wetting conditions with an
experimental set-up that delivers high contrast images of the rivulet
for quantitative analysis at both local and large scales. This is
achieved by lighting the rivulet from below through a Fresnel lens
(lateral dimenions 1.2\,m by 0.8\,m, focal length $f=1$\,m) and a
transparent substrate, and placing a camera at the point where the
light rays emanating from a light emitting diode are focussed by the
lens. The minimal distance of camera and light source is $4f$ if the
conjugated positions are chosen symmetrically, at equal distance to
the lens. In the absence of any obstacle, the substrate appears evenly
illuminated, as light rays through every point of the lens enter the
camera's entrance pupil. The set-up is very sensitive to the quality
of the optical components in play, as there is now essentially only
one light ray through any point in the substrate, at least in the
limit where the size of the light source and the objective diaphragm
(both about 1\,mm in diameter) are negligible compared to the length
of the light path ($\sim 4f = 4\,$m). In practice we see mainly
chromatic aberrations and diffraction on the rings of the Fresnel lens.

\begin{figure}
  \includegraphics[width=\textwidth]{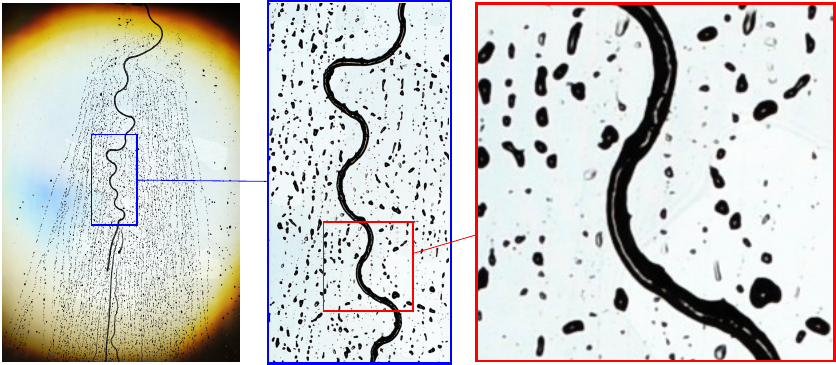}
  \caption{Photograph of the meander, with magnified details in the
    middle and to the right. Note how the white line in the rivulet,
    visible in the most zoomed detail to the right, indicates the
    locus of maximum rivulet thickness.}
\label{fig:photo}
\end{figure}

The rivulets shown in the video are distilled water running down a
vertical polyethylene or glass substrate at flow rates of a few
$\mathrm{cm}^3/\mathrm{s}$.

With a 12\,Megapixel digital camera, the resolution is sufficiently
good to measure the curvature of the meander down to the scale of the
rivulet width (see Fig.~\ref{fig:photo}). We have used this recently
to find a relation between the mean roughness of the initial straight
rivulet, and the threshold flow rate at which that rivulet starts
forming meanders\cite{couvreur2012}.

The meander dynamics shows a surprising range of time scales. While
the liquid takes only about one second to run down the incline, the
lateral motion of the rivulet is typically two orders of magnitude
smaller ($\sim$1\,cm/s). Stationary shapes of meanders are sometimes
reached only after about $10^4\,$s. In principle our highly contrasted
images, compatible with automatic detection and image processing,
should allow for detailed analysis of this long dynamics.

\end{document}